# Orientation-dependent indentation response of helium-implanted tungsten

Suchandrima Das[a⊥], Hongbing Yu[a], Edmund Tarleton[b], Felix Hofmann[a*]

[a]*Department of Engineering Science, Parks Road, Oxford UK OX1 3PJ*
[b]*Department of Materials, Parks Road, Oxford UK OX1 3PJ*

[⊥]suchandrima.das@eng.ox.ac.uk
[*]felix.hofmann@eng.ox.ac.uk

**Abstract**
A literature review of studies investigating the topography of nano-indents in ion-implanted materials reveals seemingly inconsistent observations, with report of both pile-up and sink-in. This may be due to the crystallographic orientation of the measured sample point, which is often not considered when evaluating implantation-induced changes in the deformation response. Here we explore the orientation dependence of spherical nano-indentation in pure and helium-implanted tungsten, considering grains with <001>, <110> and <111> out-of-plane orientations. Atomic force microscopy (AFM) of indents in unimplanted tungsten shows little orientation dependence. However, in the implanted material a much larger, more localised pile-up is observed for <001> grains than for <110> and <111> orientations. Based on the observations for <001> grains, we hypothesise that a large initial hardening due to helium-induced defects is followed by localised defect removal and subsequent strain softening. A crystal plasticity finite element model of the indentation process, formulated based on this hypothesis, accurately reproduces the experimentally-observed orientation-dependence of indent morphology. The results suggest that the mechanism governing the interaction of helium-induced defects with glide dislocations is orientation independent. Rather, differences in pile-up morphology are due to the relative orientations of the crystal slip systems, sample surface and spherical indenter. This highlights the importance of accounting for crystallographic orientation when probing the deformation behaviour of ion-implanted materials using nano-indentation.

**Main text**
Ion-implantation is commonly used to mimic irradiation damage in materials. This cost-effective technique allows examination of specific irradiation factors. For example helium-implantation introduces irradiation-like defects, while enabling examination of the interaction of these defects with the implanted helium [1]–[3]. Similarly self-ion implantation has been extensively employed to emulate the cascade damage caused by neutron irradiation [4], [5].
Nano-indentation is often used to gain insight into the mechanical properties of the few-micron-thick ion damaged layers. Marked changes in pile-up morphology around nano-indents have been observed in irradiated materials, even for low damage levels. A review of several studies reveals seemingly contradictory observations. For example, large pile-up around 250 nm deep Berkovich nano-indents was observed in 0.3 at.% helium-implanted W-1 at. % Re [6]. But for HT9 ferritic/martensitic steel, implanted with both helium and protons, there was no noticeable difference in indent surface profile [7]. A suppression of pile-up was noticed around indents in 2 MeV $W^+$ ion implanted W-5 wt%Ta (0.04 dpa) [8]. On the other hand $Fe^+$ implantation of Fe-12 wt%Cr lead to a distinct increase in pile-up [9].

This raises the question why implantation with helium or self-ions causes pile-up increase in some materials and suppression in others. Generally, pile-up and sink-in are, respectively, associated with low and high strain-hardening potential [10], [11]. Different annealing conditions could lead to differences in strain-hardening potential and consequently different surface profiles. Interestingly, nano-indentation in annealed single crystals has shown a strong dependence of the deformation pattern on the crystallographic orientation of the crystal [12]–[14]. This was also seen in atomistic simulations concentrating on the early stages of nano-indentation in fcc crystals [15].

In most nano-indentation studies of ion-implanted material crystallographic orientation is not considered. This raises the question whether the pile-up morphology and consequently the measured hardness may be orientation-dependent. If this is the case, then conclusions drawn from indentation of material with unknown crystal orientation could be misleading.

Here we concentrate on tungsten, the main candidate material for fusion reactor armour due to its high melting point (3422 °C) and strength at high temperature [16]–[18]. Helium-defects are known to cause substantial property changes in tungsten, e.g. increased hardness [19], lattice swelling [1], [2], [20] and reduced thermal diffusivity [21]. Recently, comparing spherical nano-indents in 0.3 at.% helium-implanted and unimplanted parts of the same tungsten <001> single crystal, we found a large, localised increase in pile-up around indents in the helium-implanted material [22]. Here, we examine whether this helium-induced change in pile-up morphology is dependent on grain orientation. For each crystal orientation, besides the indent surface morphology, the deformation field beneath specific indents is investigated using a physically-based crystal plasticity finite element (CPFE) model.

A tungsten poly-crystal (99.99% purity, ~180 µm grain size) was recrystallised at 1400 °C. The surface was mechanically ground and then polished using diamond paste. Final chemo-mechanical polishing with 0.1 µm colloidal silica suspension produced a high quality surface finish. Part of the sample was implanted with helium at the National Ion Beam Centre, University of Surrey, UK. Implantation was performed at 298 K using a 2 MeV ion accelerator and a raster scanned beam to ensure a uniform implantation dose. The implantation profile, estimated using SRIM [23] (displacement energy of 68 eV, single-layer calculation model [24]), is shown in Appendix A. Using a combination of different ion energies and fluences (Appendix B) an almost uniform helium ion concentration > 3000 appm (associated damage of ~0.24 dpa) was obtained within a ~2.8 µm thick implanted layer. Frenkel pair formation is likely to be the main damage mechanism as helium-implantation-induced recoils have predominantly low energy [2]. Little defect clustering is expected given that vacancies in tungsten have low mobility at room temperature [25]–[27].

Using electron back-scatter diffraction (EBSD), grains with near <001>, <110> and <111> orientation were selected in both the unimplanted and implanted regions of the sample.

Table *1* shows the out-of-plane orientations of the twelve selected grains and their misorientation with respect to the perfect <111>, <110> or <001> out-of-plane direction.

| Implanted part of sample | Euler Angles $(\varphi_1, \varphi, \varphi_2)$[a] | Out-of-plane orientation | Misorientation with <001> or <011> or <111> (in degrees) |
|---|---|---|---|
| (001) Grain 1 | 290.4,13.5,74.7 | [22.52,6.16,97.24] | 13.5 |
| (001) Grain 2 | 42.6,9.6,301.3 | [14.25,8.66,98.6] | 9.6 |
| (011) Grain 1 | 351.1,37.4,358.2 | [1.91,60.1,79.44] | 7.69 |
| (011) Grain 2 | 161.3,35.3,190 | [-10.03,-56.91,81.6] | 11.62 |
| (111) Grain 1 | 40.3,56.8,312.9 | [-61.3,56.96,54.76] | 2.69 |
| (111) Grain 2 | 51.3,49.7,320.7 | [-48.31,59.02,64.68] | 6.75 |
| **Unimplanted part of sample** | **Euler Angles $(\varphi_1, \varphi, \varphi_2)$** | **Out-of-plane orientation** | **Misorientation with <001> or <011> or <111> (in degrees)** |
| (001) Grain 1 | 267.1,8.1,273.9 | [-14.06,0.96,99] | 8.1 |
| (001) Grain 2 | 151,11.9,32 | [10.93,17.49,97.85] | 11.9 |
| (011) Grain 1 | 290.7,38,273.3 | [-61.46,3.54,78.8] | 7.33 |
| (011) Grain 2 | 168.6,42.3,4.7 | [5.51,67.07,73.96] | 4.22 |
| (111) Grain 1 | 186.5,46.5,323.6 | [-43.05,58.38,68.84] | 10.57 |
| (111) Grain 2 | 34.6,48.3,132.2 | [55.31,-50.15,66.52] | 6.79 |

*Table 1 - List of the out-of-plane orientation and misorientation of the twelve grains considered in this study. Misorientation is calculated with respect to the perfect <111>, <110> or <001> out-of-plane direction.*

Four 500 nm deep spherical nano-indents were made for each orientation for both implanted and unimplanted areas (MTS NanoXp, Synton ~4.2 µm radius diamond tip, 50 µm spacing between indents). Use of a spherical indenter tip removes the additional complexity of in-plane indenter orientation relative to the crystal associated with non-rotationally symmetric tips, e.g. Berkovich. SEM micrographs (Zeiss Merlin FEG SEM) of the nano-indents in <111>, <001> and <110> grains in the helium-implanted and unimplanted material are shown in Appendix C. They clearly show that consistent results are obtained across the indents for each crystal orientation.

To quantify pile-up morphology, AFM was carried out on one indent per grain orientation in the implanted and unimplanted material (Figure 1). AFM measurements were done in contact mode using a Digital Instruments Dimension 3100 AFM with Bruker CONTV-A tips (10 nm nominal tip radius). Little difference is seen between indents in different grains of the unimplanted sample (Figure 1 (d)-(f)). However, in the implanted samples there are striking orientation dependent variation in pile-up morphology (Figure 1 (j)-(l)). The <001> grain shows the characteristic large pile-up and distinct slip steps we previously observed [22].

---

[a] The Euler angle convention used is as follows: $\mathbf{Z}^{(1)}=\begin{bmatrix} \cos\varphi_1 & \sin\varphi_1 & 0 \\ -\sin\varphi_1 & \cos\varphi_1 & 0 \\ 0 & 0 & 1 \end{bmatrix}$; $\mathbf{X}=\begin{bmatrix} 1 & 0 & 0 \\ 0 & \cos\varphi & \sin\varphi \\ 0 & -\sin\varphi & \cos\varphi \end{bmatrix}$; $\mathbf{Z}^{(2)}=\begin{bmatrix} \cos\varphi_2 & \sin\varphi_2 & 0 \\ -\sin\varphi_2 & \cos\varphi_2 & 0 \\ 0 & 0 & 1 \end{bmatrix}$ and the rotation matrix $\mathbf{R} = \mathbf{Z}^{(1)} * \mathbf{X} * \mathbf{Z}^{(2)}$.

Surprisingly indents in <011> and <111> grains, though in the same polycrystalline implanted sample, show very little pile-up. Indeed comparing AFM measurements of indents in <011> and <111> grains of the unimplanted and helium-implanted material could lead to the incorrect conclusion that helium does not significantly modify the deformation behaviour of tungsten.

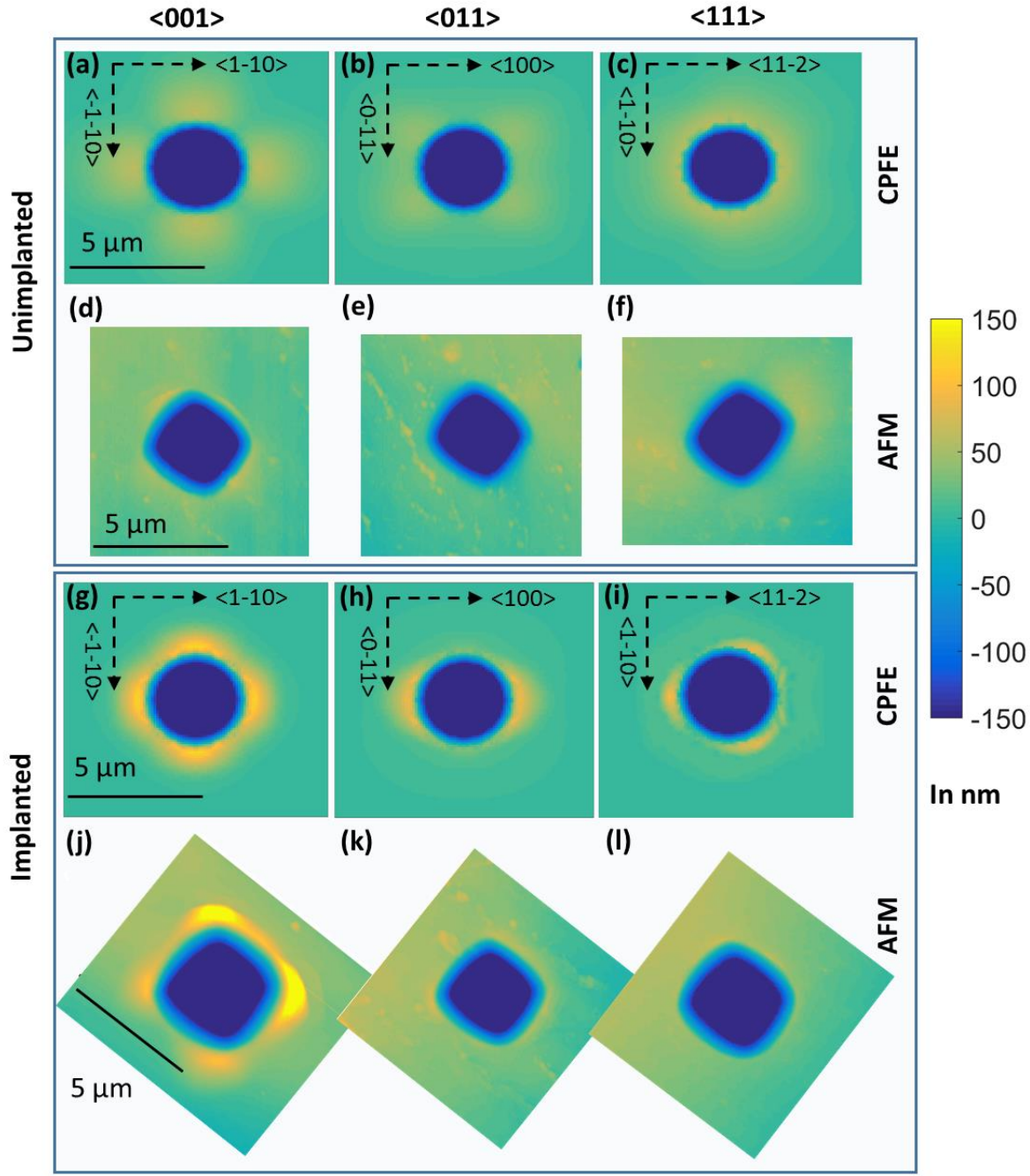

*Figure 1 - Surface profiles of residual out-of-plane displacement after indentation. CPFE simulations for the unimplanted (a) – (c) and the helium-implanted sample (g) – (i) for <001>, <011> and <111> out-of-plane crystal orientations respectively. AFM for the unimplanted (d) – (f) and the helium-implanted sample (j) – (l) for <001>, <011> and <111> out-of-plane crystal orientations respectively. The AFM micrographs have been rotated to match the in-plane orientations as labelled on the CPFE plots. The colour scale and 5 µm scale bar are the same for all plots. Although the indent depth below the surface, after unloading, is ~400 nm, a*

*colour-scale of -150 to 150 nm is used as we concentrate on investigating pile-up morphology on the sample surface.*

Orientation-dependent differences are also seen in the load-displacement curves shown in Figure 2. Each curve is the average the load-displacement response recorded from the four indents per orientation. In the unimplanted grains there is little difference between the different crystal orientations. In the helium-implanted material, on the other hand, the load for <001> orientation is ~20% higher than for <011> and ~30% higher than for <111>.

To explore the origin of the orientation-dependence of indentation behaviour in the helium-implanted material we consider a CPFE model of the indentation process. Recently we developed a CPFE formulation to simulate nano-indentation of a helium-implanted <001> oriented tungsten single crystal [28]. The formulation is based on a hypothesis derived from a comparative study of nano-indentation and micro-beam Laue diffraction measurements performed on helium-implanted and unimplanted parts of a tungsten <001> crystal [22]. Here, we use the same CPFE formulation with the appropriate orientation matrix to simulate indentation of the <011> and <111> grain of the polycrystalline tungsten sample. All other parameters of the model were kept constant. Below we briefly discuss the model and the underlying hypothesis and further details can be found elsewhere [28].

Micro-beam Laue diffraction measurements of the deformation zone beneath indents in a <001> single crystal, showed a more tightly confined plastic zone in the helium-implanted sample than in the unimplanted material [22]. Increased pile-up, slip steps and increased indentation load were observed in the implanted material. Based on these observations, we hypothesised the following mechanism for the interaction of glide dislocations and helium-implantation-induced defects (known to consist predominantly of Frenkel pairs that cannot recombine as helium occupies the vacancy [2]). We propose that initially, helium-defects strongly obstruct gliding dislocations and cause a pronounced hardening. However, with progressive deformation, passing dislocations facilitate release of helium from the defect cluster and consequently recombination of Frenkel pairs. Reduced defect density channels are thus formed that allow easier propagation of subsequent dislocations. This leads to a localisation of deformation, which in turn is the origin of the large pile-up and slip steps observed for the <001> orientation. A model based on this hypothesis was implemented in a CPFE user material subroutine (UMAT) for Abaqus where strain softening was applied to the helium-implanted layer.

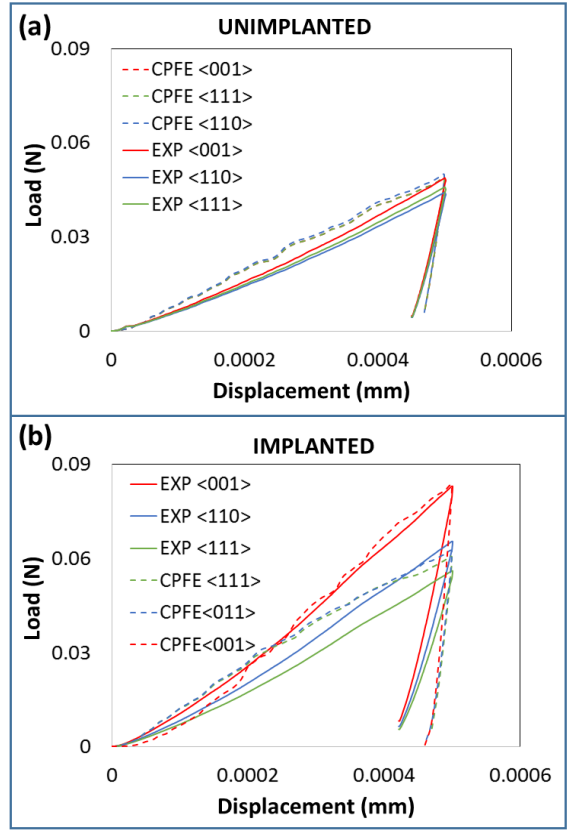

*Figure 2 – Load versus displacement curves (a) unimplanted sample and (b) helium-implanted sample for the <001>, <011> and <111> from the nano-indentation experiment and CPFE simulation superimposed. The solid curves i.e. those measured from nano-indentation experiment are the average of twelve measurements performed on four different indents (three measurements on each).*

The UMAT is based on a user-element developed by Dunne et al. [29] and is founded on the theory of multiplicative decomposition of the deformation gradient into elastic and plastic components [30-31]. Briefly, the CPFE formulation constrains slip to applicable slip-systems (assumed to be the 12 {110} slip planes with a/2<111> slip vector directions [32]). The slip rate is governed by a physically-based constitutive law that considers the glide of thermally activated dislocations in a field of pinning obstacles [29]. Taylor hardening is implemented, where the critically resolved shear stress (CRSS), with initial value $\tau_c^0$, is increased as a function of evolving densities of geometrically necessary dislocations (GNDs) [33]. The CRSS in the unimplanted sample is $\tau_c = \tau_c^0 + CG\,b^\lambda\sqrt{\rho_{GND}}$. A modified form of this equation is used in the helium-layer; $\tau_c = \tau_c^0 + CG\,b^\lambda\sqrt{\rho_{GND}} + \tau_H$ where the additional $\tau_H$ term accounts for the increased resistance to dislocation glide due to helium-defects. To account for the strain softening, $\tau_H$ is reduced at the end of each time increment, $\Delta t$, if the material point was deformed plastically. The reduction in $\tau_H$ represents the gradual formation of defect-free regions, and consequently easier dislocation glide. $\tau_H$ is considered to be a function of the total accumulated crystallographic slip and the rate at which it decreases is estimated to be proportional to its current value i.e. the current helium-defect concentration[b]. This suggests an exponential softening:

$$\tau_H^t = \tau_H^0 e^{-(\beta_p^t/\gamma)} \qquad (1)$$

$$\beta_p^{t+\Delta t} = \beta_p^t + \sum_{\lambda=1}^{n} \dot{\beta}_p^\lambda \Delta t \qquad (2)$$

where $\dot{\beta}_p^\lambda$ is the crystallographic slip rate on slip system $\lambda$, $\beta_p^t$ and $\beta_p^{t+\Delta t}$ are the total accumulated crystallographic slip, summed over all slip systems, at the start and end of the time increment, and $\tau_H^0$ is the initial value of $\tau_H$. Only three UMAT parameters were fitted to the nano-indentation and AFM results of the <001> grain: the initial unimplanted CRSS, $\tau_c^0$, the hardening prefactor $C$ and only one additional parameter for the helium-layer, the softening-rate $\gamma$. All other UMAT parameters, including $\tau_H^0$ were physically-derived, or taken from literature values [28] .

The geometry for CPFE simulations was a 3D, 20×20×20 µm³ sample block and a 4.2 µm radius spherical indenter simulated in Abaqus (Dassault Systèmes, Providence, RI, USA). Based on symmetry, for the <001> and <011> grains, the model simulated one quarter of the experimental setup. For the <111> grain, a model simulating a third of the experimental setup was used. The 20 µm high block was partitioned into two layers: a 3 µm thick top layer and 17 µm thick bottom layer. When simulating indentation on the helium-implanted tungsten, the top layer was assigned material parameters of implanted tungsten and the bottom layer that of pure tungsten. The indenter was subjected to a displacement of 0.5 µm into the sample block. A structured finite element biased mesh with >39500 20-noded quadratic hexahedral elements with reduced integration was used (C3D20R) (Appendix D); with an element size of 50 nm at the indent.

The simulated indent surface-profiles, after unloading, for both samples are shown in Figure 1. The AFM micrographs in Figure 1 are rotated to have the same in-plane orientation as the profiles predicted by CPFE. The simulated load-displacement curves for each grain orientation in both the implanted and unimplanted samples are shown superimposed on the experimental measurements in Figure 2. A good quantitative agreement is observed between CPFE and experimental results for the unimplanted sample, particularly evident in the load-displacement curves. Four, two and three-fold symmetry can be seen in the CPFE predicted surface profiles of the (001), (011) and (111) grains respectively (Figure 1 (a)-(c)).

The CPFE model reproduces the experimental pile-up for all three grains in the implanted sample remarkably well. In particular, the model captures the much lower pile-up in the <011> and the <111> grains compared to the <001> grain. In terms of the mechanical response, the <001> grain reaches ~14% higher load than the <011> and <111> grains; with the latter two producing a very similar load response. It is important to note here that the parameters of the model were determined solely based on the indentation results of the <100> oriented crystal. For the simulations of the <011> and <111> oriented grains all parameters except for the input crystal orientation were kept unchanged.

Anisotropy in indentation behaviour has been previously noted in ∝-Ti [35] and Be [36] polycrystals, where modulus and hardness decreased significantly with increasing angle of inclination between the c-axis and the indentation axis. Orientation-dependent mechanical

---

[b] $\left.\frac{\partial \tau_H}{\partial \beta_p}\right|_{t+\Delta t} = -\tau_H/\gamma$

performance has been seen in tungsten too, where the <001> single crystal was found to be the best performing kinetic energy penetrator, owing to favourable slip during loading and shear localization [37]. It is interesting that orientation-dependent differences in indentation response are as pronounced at the nano-scale. The results highlight the importance of determining the crystallographic orientation of the ion-implanted sample for accurate evaluation of the ion-induced alteration in deformation behaviour.

Quantitative agreement between CPFE predictions and experimental results for the implanted sample inspires some confidence in the hypothesis that localised deformation through slip channels, formed by dislocations interacting with helium-defects, can cause the large pile-up and increased hardening. In the 1960s' Makin et al. proposed a mathematical theory linking the creation of large pile-up and slip-steps to an accelerating fall in the force resisting dislocation glide [38]. As the defect reduction rate will depend on the number of defects, an exponential decrease in $\tau_H$ as implemented in the CPFE formulation in Eq. (2), is reasonable. The model uniquely captures the four-fold pile-up increase in the <001> grain in the implanted samples, affirming the strain-softening hypothesis. Strain-softening and consequent slip channel formation in irradiated materials, both fcc and bcc, has been experimentally observed in numerous studies [39]–[41]. This is of particular concern as it may lead to untimely failure due to loss of ductility.

The CPFE predictions also provide further insight into the underlying deformation zone beneath indents. Comparison of the field of effective plastic strain beneath indents in each grain for both samples (Figure 3) shows an orientation-independent trend. The deformation field becomes more confined near the indent in all grains in the implanted material, compared to a more widespread deformation zone in the unimplanted material.

Put together, these results, suggest that the underlying mechanism responsible for the modified behaviour of the implanted material, i.e. the interaction of dislocation with helium-induced defects, is orientation-independent. The changes in the pile-up morphology for different crystal orientations are simply a product of the relative orientations of the active slip systems, the indenter and the sample surface.

In summary, we have examined the effect of helium-implantation in grains of three different orientations. It was found that the (001)-oriented grains showed ~70% increased hardness and ~172% increased pile-up compared to unimplanted tungsten. In contrast, the (011) and (111) grains show negligible change in pile-up and only a ~30% increase in hardness compared to the unimplanted sample. CPFE based on the application of strain-softening in the helium-implanted layer, was able to reproduce the experimental results for all three grains with surprising accuracy. The fact that significantly different indentation behaviour is observed for different grain orientations in the implanted material highlights the importance of considering crystal orientation when interpreting nano-indentation data.

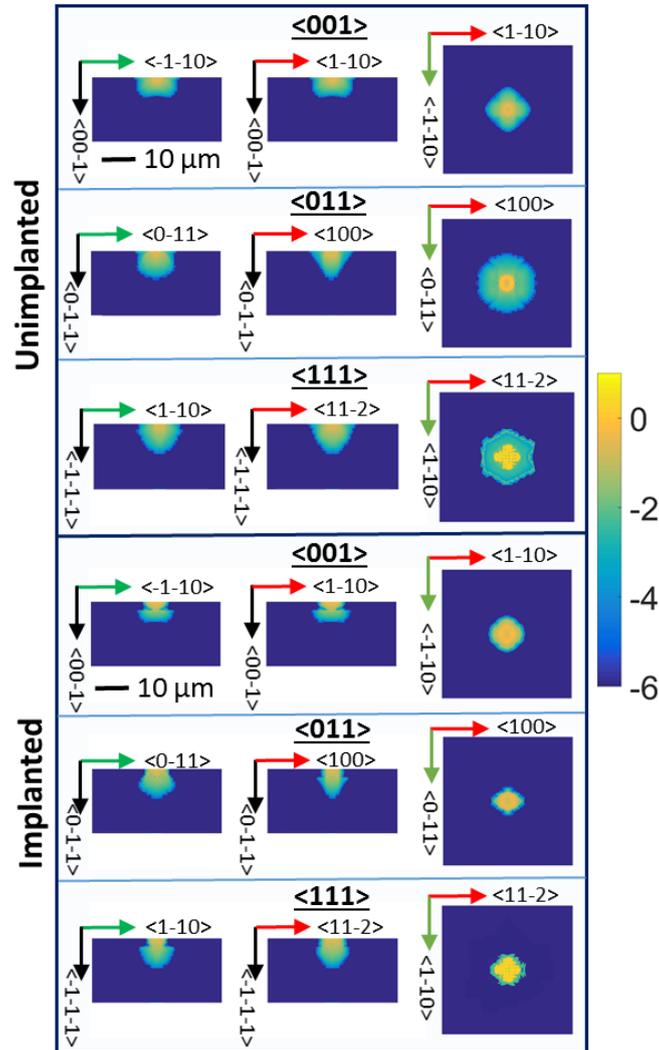

*Figure 3 – A comparison of the effective plastic strain (in logarithmic scale) beneath indents in all three grains for both the unimplanted and the helium-implanted sample as predicted by the CPFE simulation. In each case, the strain is shown on the YZ, XZ and the XY cross-sections from left to right.*

**ACKNOWLEDGEMENTS**
We thank Prof. D.E.J. Armstrong for providing the samples, and Dr. N. Peng for performing the ion-implantation. This work was funded by Leverhulme Trust Research Project Grant RPG-2016-190. Ion implantations were performed under the UK Engineering and Physical Sciences Research Council grant EP/H018921/1. ET acknowledges financial support from the Engineering and Physical Sciences Research Council under Fellowship grant EP/N007239/1

## Appendix A

Figure A.1 shows the helium-implantation profile as performed on part of the tungsten polycrystalline sample at room temperature.

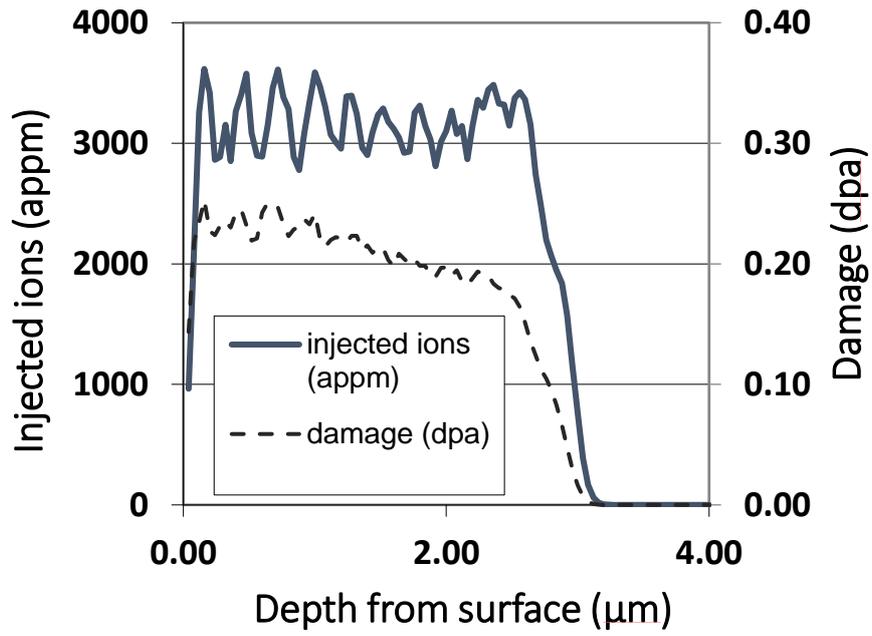

*Figure A.1 - (a) Helium-implantation profile predicted by SRIM for the ion-implanted part of the tungsten sample.*

## Appendix B

Table B.1 shows the different ion energies and fluences used for helium-implantation.

| Ion Energy (MeV) | Fluence ($\times 10^{15}$ ions/cm²) |
|---|---|
| 0.05 | 2.40 |
| 0.1 | 1.80 |
| 0.2 | 4.20 |
| 0.3 | 1.20 |
| 0.4 | 4.80 |
| 0.6 | 5.20 |
| 0.8 | 5.00 |
| 1 | 5.00 |
| 1.2 | 5.00 |
| 1.4 | 5.00 |
| 1.6 | 5.50 |
| 1.8 | 7.00 |
| 2 | 5.00 |

*Table B.1 – List of 12 ion energies and fluences used for helium-implantation.*

## Appendix C

### **SEM micrographs of Indents**

Figure C.1 and Figure C.2 show SEM micrographs of the spherical indents made in <001>, <011> and <111> grains in the unimplanted and helium-implanted sample respectively. Four indents were made for each orientation in each sample (with 50 µm spacing between indents for indents in the same grain). SEM micrographs of the indent topography show consistent results across each orientation and implantation condition.

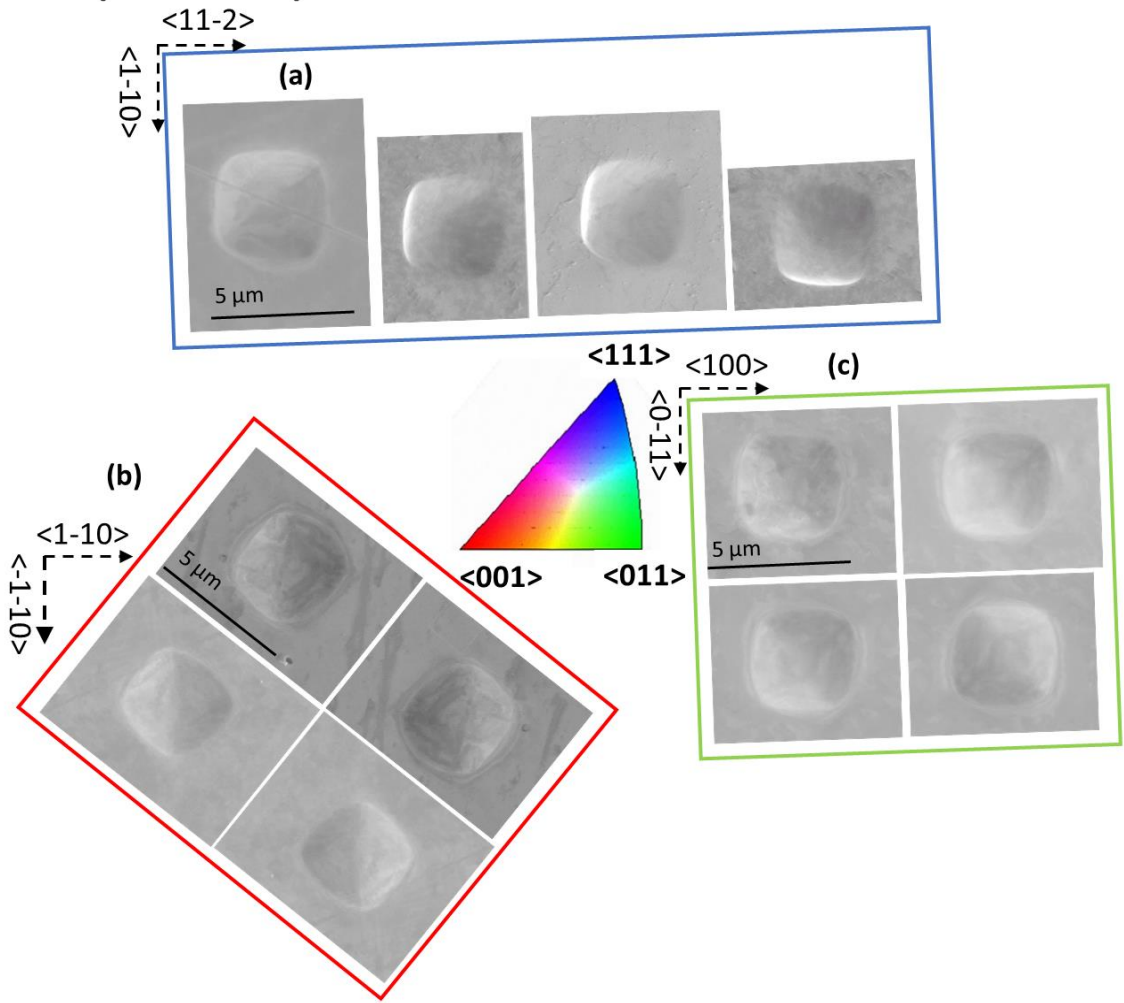

*Figure C.1 – SEM micrographs of four indents in (a) <001>, (b) <011> and (c) <111> grains in the unimplanted sample; The SEM micrographs have been rotated to match the in-plane crystal orientation as indicated by dotted black labelled arrows.*

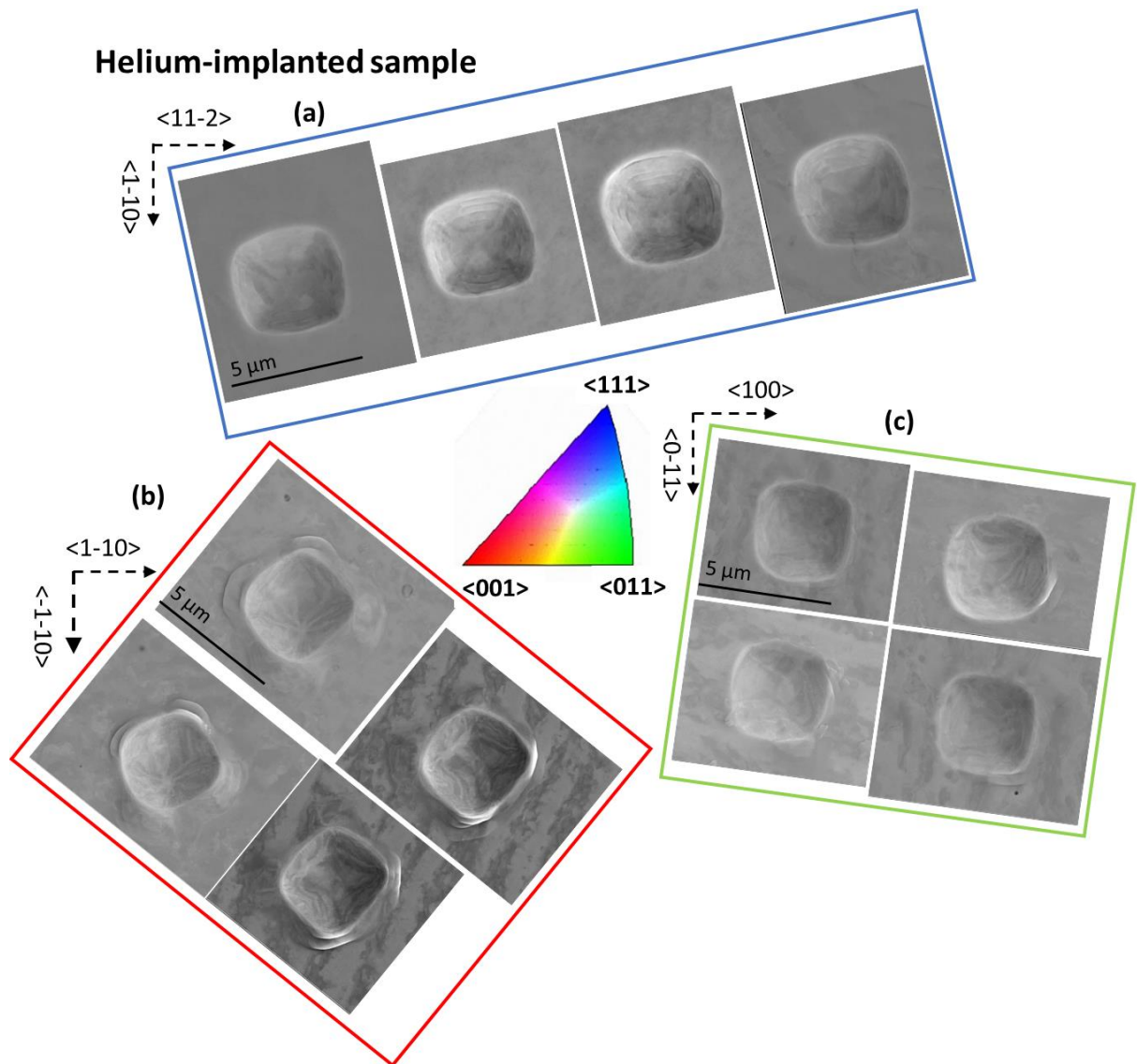

*Figure C.2 - SEM micrographs of four indents in (a) <001>, (b) <011> and (c) <111> grains in the helium-implanted sample; The SEM micrographs have been rotated to match the in-plane crystal orientation as indicated by dotted black labelled arrows.*

Appendix D

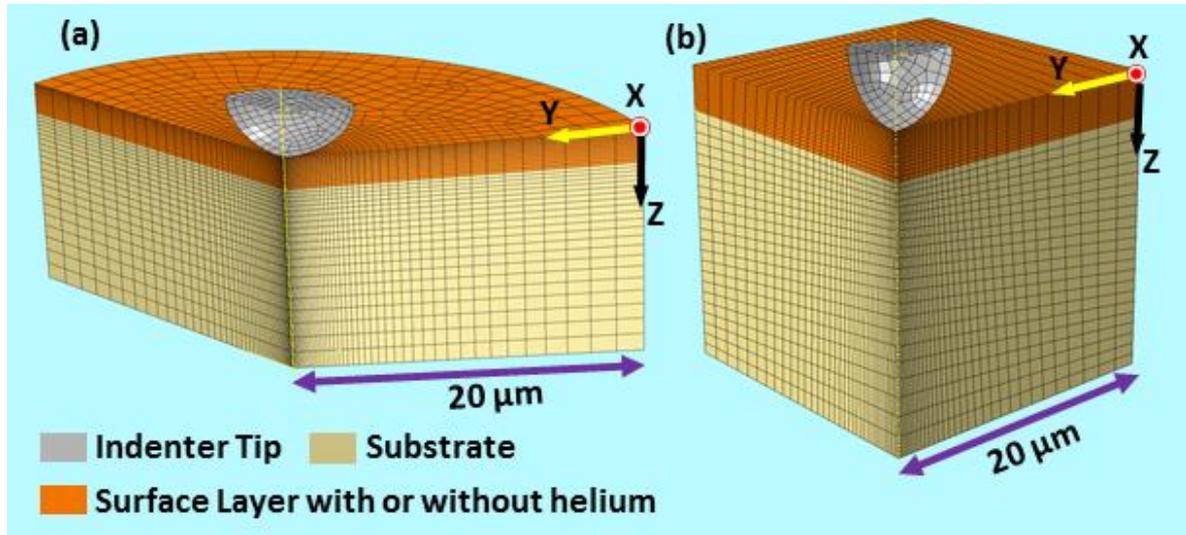

*Figure D.1 – Rendering of the finite element mesh used for 3D crystal plasticity simulations of the tungsten sample indented by a 4.2 µm radius spherical indenter (a) one-third model for the <111> orientation and (b) quarter model for the <001> and <011> orientations. The X, Y, Z coordinate frame used throughout this work is superimposed.*

**Details of the CPFE formulation**
Details about the formulation can be found elsewhere [28]. The values for the material properties used in the formulation is provided below in Table D.1.

| Material Property | Value | Reference |
|---|---|---|
| Elastic modulus $E$ | 410 GPa | [42]–[44], [46] |
| Shear modulus $G$ | 164.4 GPa | [42]–[44], [46] |
| Poisson's ratio $v$ | 0.28 | [42]–[44], [46] |
| Burgers' vector $b$ | $2.7 \times 10^{-10}$ m | [47] |
| Stress boundary conditions $\sigma_{xx}^{BC} = \sigma_{yy}^{BC}$ | -262.704 MPa | Details explained in section 1 |
| $\tau_H^0$ | 750 MPa | [28] |
| Helmholtz free energy $\Delta F$ | 0.85 eV | [28] |
| Boltzmann constant $k$ | $1.381 \times 10^{-23}$ J/K | [48] |
| Temperature $T$ | 298 K | Room temperature assumed similar to experimental conditions |
| Attempt frequency $v$ | $1 \times 10^{19}$ s$^{-1}$ | [49] |
| Density of statistically stored dislocations, $\rho_{SSD}$ | $1 \times 10^{10}$ m$^{-2}$ | [28] |
| Density of mobile dislocations $\rho_m$ | $1.75 \times 10^{13}$ m$^{-2}$ | [28] |
| Probability of pinning $\Psi$ | $0.657 \times 10^{-2}$ | Value chosen and kept fixed |
| $\tau_c^0$ | 360 MPa | Fitted to experimental data of unimplanted sample |
| $\gamma$ | 0.025 | Fitted to experimental data of helium-implanted sample |
| $C'$ | 0.0065 | Fitted to experimental data of unimplanted sample |

*Table D.1- List of parameters used in the constitutive law in the CPFE formulation and their corresponding values.*

1. Applied Boundary Conditions

The boundary conditions applied to the indentation model in Abaqus included symmetric XZ and YZ planes, a traction free top surface, and fixed displacement and rotation boundary conditions on the remaining surfaces. An additional boundary condition was applied to the helium-implanted layer to account for helium-implantation induced residual stresses. Detailed description of the method of calculation of residual stresses generated by helium-implantation can be found elsewhere [1]. Briefly, the residual stresses can be expressed as a function of the out-of-plane lattice swelling i.e. $\varepsilon_{zz}$ induced by helium

$$\sigma_{xx}^r = \frac{-E\varepsilon_{zz}}{3(1+v)} = \sigma_{yy}^r \qquad (D.1).$$

In a prior study, on a 001-single crystal tungsten implanted with helium under similar conditions, $\varepsilon_{zz}^{dev}$, the deviatoric component of the out-of-plane strain, was measured by white-beam Laue diffraction [22]. Here $\varepsilon_{zz}$ is estimated from this prior obtained $\varepsilon_{zz}^{dev}$ ; $\varepsilon_{zz} = 3/2 \, \varepsilon_{zz}^{dev}$ . It is assumed that $\varepsilon_{xx}$ and $\varepsilon_{yy}$ component of the total strain tensor is zero, (deformation along X and Y directions are restricted in order to maintain geometrical continuity between the implanted layer and the substrate). Knowing $\varepsilon_{zz}^{dev}$ in the helium layer to be ~ 530 $\times 10^{-6}$, $\sigma_{xx}^r = \sigma_{yy}^r$ are computed to be -260 MPa.

2. Scaling with effective modulus

The sample block was assigned elastic properties of tungsten i.e. elastic modulus of 410 GPa and Poisson's ration of 0.28 [42]–[44]. The indenter, designed as a discrete rigid wire frame was not assigned any material properties. This was done to avoid a full meshing and to allow increased simulation size. The material properties of the diamond indenter tip (with modulus of $E_i = 1143$ GPa) used in the experiment, was accounted for in the simulation by scaling the results with an effective modulus $E_{eff}$ (322.58 GPa), where, $\frac{1}{E_{eff}} = \frac{1-\nu_W^2}{E_W} + \frac{1-\nu_i^2}{E_i}$ [45].

$$\tau_c = \tau_c^0 + CG\, b^\lambda \sqrt{\rho_{GND}} + \tau_H \qquad (D.2)$$

**REFERENCES**
[1] S. Das, W. Liu, R. Xu, and F. Hofmann, "Helium-implantation-induced lattice strains and defects in tungsten probed by X-ray micro-diffraction," *Mater. Des.*, vol. 160, pp. 1226–1237, 2018.
[2] F. Hofmann *et al.*, "Lattice swelling and modulus change in a helium-implanted tungsten alloy: X-ray micro-diffraction, surface acoustic wave measurements, and multiscale modelling," *Acta Mater.*, vol. 89, pp. 352–363, 2015.
[3] I. DeBroglie, C. E. Beck, W. Liu, and F. Hofmann, "Temperature Dependence of Helium--Implantation--Induced Lattice Swelling in Polycrystalline Tungsten: X--ray Micro--Diffraction and Eigenstrain Modelling," *Scr. Mater.*, vol. 107, p. 4, 2015.
[4] X. Yi, M. L. Jenkins, M. A. Kirk, Z. Zhou, and S. G. Roberts, "In-situ TEM studies of 150 keV W+ ion irradiated W and W-alloys: Damage production and microstructural evolution," *Acta Mater.*, vol. 112, pp. 105–120, 2016.
[5] D. E. J. Armstrong, X. Yi, E. A. Marquis, and S. G. Roberts, "Hardening of self ion implanted tungsten and tungsten 5-wt% rhenium," *J. Nucl. Mater.*, vol. 432, no. 1–3, pp. 428–436, 2013.
[6] C. E. Beck *et al.*, "Correcting for contact area changes in nanoindentation using surface acoustic waves," *Scr. Mater.*, vol. 128, pp. 83–86, 2017.
[7] P. Hosemann *et al.*, "Nanoindentation on ion irradiated steels," *J. Nucl. Mater.*, vol. 389, no. 2, pp. 239–247, 2009.
[8] D. E. J. Armstrong, A. J. Wilkinson, and S. G. Roberts, "Mechanical properties of ion-implanted tungsten–5 wt% tantalum," *Phys. Scr.*, vol. 2011, no. T145, p. 14076, 2011.
[9] C. D. Hardie, S. G. Roberts, and A. J. Bushby, "Understanding the effects of ion irradiation using nanoindentation techniques," *J. Nucl. Mater.*, vol. 462, pp. 391–401, 2015.
[10] W. D. Nix, "Elastic and plastic properties of thin films on substrates: nanoindentation techniques," *Mater. Sci. Eng. A*, vol. 234–236, pp. 37–44, 1997.
[11] C. Oliver and M. Pharr, "An improved technique for determining hardness and elastic modulus using load and displacement sensing indentation experiments," *Journal of Materials Research*, vol. 7, no. 11. pp. 1564–1583, Jun-1992.
[12] Y. Gaillard, C. Tromas, and J. Woirgard, "Study of the dislocation structure involved in a nanoindentation test by atomic force microscopy and controlled chemical etching," *Acta Mater.*, vol. 51, no. 4, pp. 1059–1065, 2003.
[13] D. F. Bahr, D. E. Kramer, and W. W. Gerberich, "Non-linear deformation mechanisms during nanoindentation," *Acta Mater.*, vol. 46, no. 10, pp. 3605–3617, 1998.
[14] M. Hollatz, M. Bobeth, W. Pompe, and V. Marx, "Orientation dependent crack patterns in alumina films on NiAl single crystals due to spherical indentation," *Acta Mater.*, vol. 44, no. 10, pp. 4149–4159, 1996.
[15] J. Li, K. J. Van Vliet, T. Zhu, S. Yip, and S. Suresh, "Atomistic mechanisms governing elastic limit and incipient plasticity in crystals," *Nature*, vol. 418, p. 307, Jul. 2002.
[16] M. Rieth *et al.*, "Recent progress in research on tungsten materials for nuclear fusion applications in Europe," *J. Nucl. Mater.*, vol. 432, no. 1, pp. 482–500, 2013.
[17] M. Ekman, K. Persson, and G. Grimvall, "Phase diagram and lattice instability in tungsten ± rhenium alloys," *J. Nucl. Mater.*, vol. 278, pp. 276–279, 2000.
[18] N. Wei, T. Jia, X. Zhang, T. Liu, Z. Zeng, and X. Yang, "First-principles study of the phase stability and the mechanical properties of W-Ta and W-Re alloys," *AIP Adv.*, vol. 4, no. 5, p. 57103, May 2014.

[19] D. E. J. Armstrong, P. D. Edmondson, and S. G. Roberts, "Effects of sequential tungsten and helium ion implantation on nano-indentation hardness of tungsten," *Appl. Phys. Lett.*, vol. 251901, no. 25, pp. 1–5, 2013.

[20] I. De Broglie, C. E. Beck, W. Liu, and F. Hofmann, "Temperature dependence of helium-implantation-induced lattice swelling in polycrystalline tungsten: X-ray micro-diffraction and Eigenstrain modelling," *Scr. Mater.*, vol. 107, 2015.

[21] F. Hofmann, D. R. Mason, J. K. Eliason, A. A. Maznev, K. A. Nelson, and S. L. Dudarev, "Non-Contact Measurement of Thermal Diffusivity in Ion-Implanted Nuclear Materials," *Sci. Rep.*, vol. 5, p. 16042, 2015.

[22] S. Das, D. E. J. Armstrong, Y. Zayachuk, W. Liu, R. Xu, and F. Hofmann, "The effect of helium implantation on the deformation behaviour of tungsten: X-ray micro-diffraction and nanoindentation," *Scr. Mater.*, vol. 146, pp. 335–339, Mar. 2018.

[23] J. F. Ziegler and J. Biersack, "SRIM – The stopping and range of ions in matter (2010)," *Nucl. Inst. Methods Phys. Res. B*, vol. 268, pp. 1818–1823, 2010.

[24] "ASTM E521–96(2009) e1, Standard Practice for Neutron Radiation Damage Simulation by Charged-Particle Irradiation, ASTM International, West Conshohocken, PA, 2009, USA."

[25] A. Debelle, M. F. F. Barthe, and T. Sauvage, "First temperature stage evolution of irradiation-induced defects in tungsten studied by positron annihilation spectroscopy," *J. Nucl. Mater.*, vol. 376, no. 2, pp. 216–221, May 2008.

[26] P. M. Derlet, D. Nguyen-Manh, and S. L. Dudarev, "Multiscale modeling of crowdion and vacancy defects in body-centered-cubic transition metals," *Phys. Rev. B - Condens. Matter Mater. Phys.*, vol. 76, no. 5, 2007.

[27] D. Nguyen-Manh, A. P. Horsfield, and S. L. Dudarev, "Self-interstitial atom defects in bcc transition metals: Group-specific trends," *Phys. Rev. B - Condens. Matter Mater. Phys.*, vol. 73, no. 2, 2006.

[28] S. Das, H. Yu, E. Tarleton, and F. Hofmann, "Hardening and Strain Localisation in Helium-Ion-Implanted Tungsten," *arXiv Prepr. arXiv1901.00745*, Dec. 2018.

[29] F. P. E. Dunne, D. Rugg, and A. Walker, "Lengthscale-dependent, elastically anisotropic, physically-based hcp crystal plasticity: Application to cold-dwell fatigue in Ti alloys," *Int. J. Plast.*, vol. 23, no. 6, pp. 1061–1083, 2007.

[30] E. H. Lee, "Elastic-Plastic Deformation at Finite Strains," *J. Appl. Mech.*, vol. 36, no. 1, pp. 1–6, Mar. 1969.

[31] S. Das, F. Hofmann, and E. Tarleton, "Consistent determination of geometrically necessary dislocation density from simulations and experiments," *Int. J. Plast.*, vol. 109, pp. 18–42, May 2018.

[32] C. Marichal, H. Van Swygenhoven, S. Van Petegem, and C. Borca, "{110} Slip with {112} slip traces in bcc Tungsten.," *Sci. Rep.*, vol. 3, no. 211, p. 2547, 2013.

[33] G. I. Taylor, "The Mechanism of Plastic Deformation of Crystals. Part I. Comparison with Observations," *Proc. R. Soc. A Math. Phys. Eng. Sci.*, vol. 145, no. 855, pp. 388–404, 1934.

[34] Y. Wang, D. Raabe, C. Klüber, and F. Roters, "Orientation dependence of nanoindentation pile-up patterns and of nanoindentation microtextures in copper single crystals," *Acta Mater.*, vol. 52, no. 8, pp. 2229–2238, 2004.

[35] T. B. Britton, H. Liang, F. P. E. Dunne, and A. J. Wilkinson, "The effect of crystal orientation on the indentation response of commercially pure titanium: experiments and simulations," *Proc. R. Soc. A Math. Phys. Eng. Sci.*, vol. 466, no. 2115, pp. 695–719, Mar. 2010.

[36] V. Kuksenko, S. Roberts, and E. Tarleton, "The hardness and modulus of polycrystalline beryllium from nano-indentation," *Int. J. Plast.*, vol. 116, pp. 62–80, 2019.

[37] P. W. . Bruchey, W.J; Horwath, E.J.;Kingman, "Orientation dependence of deformation and penetration behavior of tungsten single-crystal rods," no. 23061891, 1991.

[38] M. J. Makin and J. V. Sharp, "A Model of Lattice Hardening in Irradiated Copper Crystals with the External Characteristics of Source Hardening," *Phys. status solidi*, vol. 9, no. 1, pp. 109–118, 1965.

[39] A. J. Bushby, S. G. Roberts, and C. D. Hardie, "Nanoindentation investigation of ion-irradiated Fe-Cr alloys using spherical indenters," *J. Mater. Res.*, vol. 27, no. 1, pp. 85–90, 2012.

[40] M. Victoria *et al.*, "The microstructure and associated tensile properties of irradiated fcc and bcc metals," *J. Nucl. Mater.*, vol. 276, no. 1, pp. 114–122, 2000.

[41] Y. Dai and M. Victoria, "Defect Cluster Structure and Tensile Properties of Copper Single Crystals Irradiated With 600 MeV Protons," *MRS Proc.*, vol. 439, p. 319, 1996.

[42] F. H. Featherston and J. R. Neighbours, "Elastic constants of tantalum, tungsten, and molybdenum," *Phys. Rev.*, vol. 130, no. 4, pp. 1324–1333, 1963.

[43] D. I. Bolef and J. De Klerk, "Elastic Constants of Single-Crystal Mo and W between 77° and 500°K," *J. Appl. Phys.*, vol. 33, no. 7, pp. 2311–2314, Jul. 1962.

[44] C. A. Klein and G. F. Cardinale, "Young's modulus and Poisson's ratio of CVD diamond," *Diam. Relat. Mater.*, vol. 2, pp. 918–923, 1993.

[45] L. Ma, L. Levine, R. Dixson, D. Smith, and D. Bahr, "Effect of the Spherical Indenter Tip Assumption on the Initial Plastic Yield Stress," *Nanoindentation Mater. Sci.*, vol. 22, no. 6, pp. 1656–1661, 2012.

[46] R. A. Ayres *et al.*, "Elastic constants of tungsten−rhenium alloys from 77 to 298 °K," *J. Appl. Phys.*, vol. 46, no. 4, pp. 1526–1530, Apr. 1975.

[47] B. N. Dutta and B. Dayal, "Lattice Constants and Thermal Expansion of Palladium and Tungsten up to 878 C by X-Ray Method," *Phys. status solidi*, vol. 3, no. 12, pp. 2253–2259, Jan. 1963.

[48] C. A. Sweeney, W. Vorster, S. B. Leen, E. Sakurada, P. E. McHugh, and F. P. E. Dunne, "The role of elastic anisotropy, length scale and crystallographic slip in fatigue crack nucleation," *J. Mech. Phys. Solids*, vol. 61, no. 5, pp. 1224–1240, May 2013.

[49] A. Cottrell, *An introduction to metallurgy*. Universities Press, 1990.

[50] C. C. Fu and F. Willaime, "Ab initio study of helium in α-Fe: Dissolution, migration, and clustering with vacancies," *Phys. Rev. B - Condens. Matter Mater. Phys.*, vol. 72, no. 6, p. 64117, Aug. 2005.

[51] L. Sandoval, D. Perez, B. P. Uberuaga, and A. F. Voter, "Competing kinetics and he bubble morphology in W," *Phys. Rev. Lett.*, vol. 114, no. 10, 2015.

[52] J. C. Slater, "Atomic radii in crystals," *J. Chem. Phys.*, vol. 41, no. 10, pp. 3199–3204, 1964.

[53] D. Cereceda, J. Marian, and J. Marian, "Nudged Elastic Band Simulations of Kink Pairs in Tungsten," 2015.

[54] C. R. Weinberger, B. L. Boyce, and C. C. Battaile, "Slip planes in bcc transition metals," *Int. Mater. Rev.*, vol. 58, no. 5, pp. 296–314, 2013.

[55] G. Po *et al.*, "A phenomenological dislocation mobility law for bcc metals," *Acta Mater.*, vol. 119, pp. 123–135, 2016.

[56] G. D. Samolyuk, Y. N. Osetsky, and R. E. Stoller, "The influence of transition metal solutes on the dislocation core structure and values of the Peierls stress and barrier in tungsten," *J. Phys. Condens. Matter*, vol. 25, no. 2, 2013.

[57] "NIST Chemistry Web book, SRD69," *National Institute of Standards and Technology*. [Online]. Available: https://webbook.nist.gov/cgi/cbook.cgi?ID=C7440337&Mask=2.